\documentclass[conf, hidelinks]{new-aiaa}
\usepackage[utf8]{inputenc}

\usepackage{graphicx}
\usepackage{upgreek}
\usepackage{dsfont}
\usepackage{mathtools}
\usepackage{siunitx}
\usepackage{longtable,tabularx}

\usepackage[usenames, dvipsnames, svgnames]{xcolor}
\usepackage{tikz}
  \usetikzlibrary{arrows}
    \tikzset{>=stealth}
  \usetikzlibrary{backgrounds}
  \usetikzlibrary{external}
    \tikzset{external/only named}
    \tikzset{external/mode=list and make}
    \tikzset{external/prefix=fig/}
    \tikzset{external/system call=%
        {lualatex \tikzexternalcheckshellescape -synctex=1 %
        -interaction=batchmode -jobname "\image" "\texsource"}%
    }
    \pgfkeys{/pgf/images/include external/.code={%
        \IfFileExists{#1.pdf}{%
          \includegraphics{{#1}.pdf}%
        }{%
          \includegraphics[draft]{{#1}.pdf}%
        }%
      }
    }
  \usetikzlibrary{math}
  \usetikzlibrary{plotmarks}
  \usetikzlibrary{positioning}

\usepackage{pgfplots}
  \usepgfplotslibrary{groupplots}
  \usepgfplotslibrary{statistics}
  \pgfplotsset{compat=newest}
  \pgfplotsset{
    scaled ticks=false,
    clip marker paths=true,
    xlabel near ticks,
    ylabel near ticks,
    tick label style={font=\footnotesize},
    label style={font=\footnotesize},
    every axis title shift=1.4pt,
  }

\usepackage{booktabs,colortbl}
\usepackage{pgfplotstable}
  \pgfplotstableset{
    every odd  row/.style={before row={\rowcolor[gray]{0.9}}},
    every head row/.style={before row=\toprule,after row=\midrule},
    every last row/.style={after row=\bottomrule},
  }

\usepackage{algorithm}
\usepackage{algorithmic}

\usepackage{hyperref}

\newcommand{\npar}{{n_{\uptheta}}} 
\newcommand{\nx}{{n_{\mathrm{x}}}} 
\newcommand{\ninp}{{n_{\mathrm{u}}}} 
\newcommand{\ny}{{n_{\mathrm{y}}}} 
\newcommand{\reals}{{\mathds{R}}} 
\newcommand{\dd}{{\mathrm{d}}} 
\newcommand{\opt}{^{*}} 
\newcommand{\trans}{^{\mathsf{T}}} 
\newcommand{\inv}{^{-1}} 
\newcommand{\normald}{\mathscr{N}} 
\newcommand{\kprev}{_{k-1}} 
\newcommand{\kpath}{_{0:N}} 
\newcommand{\elbo}{\mathcal{L}} 
\newcommand{\qspace}{\mathcal{Q}} 
\newcommand{\marg}{_{\operatorname{marg}}} 
\newcommand{\cond}{_{\operatorname{cond}}} 
\newcommand{\cross}{_{\operatorname{cross}}} 
\newcommand{\sigp}{^{(i)}} 
\DeclareMathOperator{\ident}{I} 
\DeclareMathOperator{\E}{E} 
\DeclareMathOperator{\kld}{KL} 
\DeclareMathOperator{\entro}{h} 

\title{Variational System Identification of Aircraft}

\author{Dimas Abreu Archanjo Dutra\footnote{Teaching Assistant Professor, Department of Mechanical, Materials, and Aerospace Engineering, 1306 Evansdale Drive.}}
\affil{West Virginia University, Morgantown, West Virginia, 26506-6106}

\begin{document}

\maketitle

\begin{abstract}
Variational system identification is a new formulation of maximum likelihood for estimation of parameters of dynamical systems subject to process and measurement noise, such as aircraft flying in turbulence.
This formulation is an alternative to the filter-error method that circumvents the solution of a Riccati equation and does not have problems with unstable predictors.
In this paper, variational system identification is demonstrated for estimating aircraft parameters from real flight-test data.
The results show that, in real applications of practical interest, it has better convergence properties than the filter-error method, reaching the optimum even when null initial guesses are used for all parameters and decision variables.
This paper also presents the theory behind the method and practical recommendations for its use.
\end{abstract}

\section*{Nomenclature}
\subsection*{Mathematical Symbols}
{\renewcommand\arraystretch{1.0}
\noindent\begin{longtable*}{@{}l @{\quad=\quad} l@{}}
$\reals$ & Set of real numbers \\
$\delta(t)$ & Dirac's delta \\
$\delta_{ij}$ & Kronecker's delta \\
$\ident$ & Identity matrix \\
$\dd$ & Differential operator \\
$\E[\cdot]$ & Expectation operator \\
$\normald(x;\mu, \Sigma)$ & Multivariate normal density at $x$\\
$\entro_q[X]$ & Differential entropy of $X$ under density $q$ \\
$t$ & Time \\
$T$ & Sampling period \\
$k$ & Sample index \\
$\nx$ & Number of states\\
$\ninp$ & Number of external inputs\\
$\ny$ & Number of measured outputs\\
$\npar$ & Number of unknown parameters\\
$N+1$ & Number of time samples\\
$x(t)$ & State vector \\
$u(t)$ & External input vector \\
$y_k$ & Output vector \\
$v_k$ & Measuremente noise \\
$W(t)$ & $\nx$-dimensional Wiener process \\
$w(t)$ & weak derivative of $W(t)$ \\
$\theta$  & Unknown parameters \\
$f(x,u,\theta)$ & SDE drift \\
$G(\theta)$ & SDE diffusion matrix \\
$h(x,u,\theta)$ & Output function \\
$R(\theta)$ & Measurement noise covariance \\
$Q(\theta)$ & Process noise covariance \\
$q(x_{0:N})$ & Assumed density \\
$\qspace$ & Search space of assumed densities $q$ \\
$p_\theta(a|b)$ & Probability density of the random variable $A$ at the value $a$, given $B=b$, under the parameter $\theta$ \\
$\operatorname{const}$ & Constant terms which do not change the location of maxima\\
\end{longtable*}}

\subsection*{Superscripts}
{\renewcommand\arraystretch{1.0}
\noindent\begin{longtable*}{@{}l @{\quad=\quad} l@{}}
$A\trans$ & $A$ transpose\\
$A\inv$ & $A$ inverse \\
\end{longtable*}}

\subsection*{Subscripts}
{\renewcommand\arraystretch{1.0}
\noindent\begin{longtable*}{@{}l @{\quad=\quad} l@{}}
$x_k,u_k$ & $x(kT),u(kT)$ evaluation at a sample index \\
$x_{i:j}$ & $x_i,x_{i+1},\dots,x_{j-1},x_j$ sequence range \\
\end{longtable*}}

\subsection*{Acronyms}
{\renewcommand\arraystretch{1.0}
\noindent\begin{longtable*}{@{}l @{\quad} l@{}}
DLR & \emph{Deutsches Zentrum für Luft- und Raumfahrt} \\
ELBO & Evidence Lower Bound \\
FEM & Filter-Error Method \\
KLD & Kullback--Leibler Divergence \\
MAP & Maximum \emph{a Posteriori} \\
ML & Maximum Likelihood \\
OEM & Output Error Method \\
SDE & Stochastic Differential Equation \\
VI & Variational Inference \\
\end{longtable*}}
\section{Introduction}
\lettrine{A}{ircraft} system identification, the art and science \cite{jategaonkar_flight_2005} of determining mathematical models of aircraft from flight-test data, has become increasingly important in the development of new aircraft \cite{grauer_introduction_2023}.
As the field evolves \cite{jategaonkar_aerodynamic_2004, wang_retrospective_2004, morelli_application_2005, deiler_retrospective_2023, morelli_advances_2023} the problems that can be solved at the state-of-the-art are often limited by what estimation methods can be run at the computational hardware available.

Maximum likelihood (ML) approaches such as the equation-error, output-error, and filter-error methods are widely used in aircraft system identification due to their assymptotic efficiency properties.
When both process noise and measurement noise are considered, the ML estimator is the filter-error method (FEM), which employs a Kalman filter to approximate the prior predictive distribution of each measurement, given the previous measurements up to that point.
There are multiple formulations of this method, but they either require solution of a Riccati equation, the most common approach in the aircraft literature \cite[Chap.~5]{maine_formulation_1981, grauer_new_2015, jategaonkar_flight_2015}; or they require the estimation of the filter gain as an independent parameter  \cite{wills_gradient-based_2008}, the most common approach for black-box estimation, which can lead to unstable predictors in the optimization process.
A consequence of these limitations is that the FEM requires good initial parameter estimates for convergence to ocurr.

Recently \cite{courts_variational_2021, courts_variational_2023}, a new formulation of the ML estimator has been proposed for systems subject to process and measurement noise, using variational inference (VI).
This approach was then reframed as unconstrained optimization and reparameterized for steady-state problems by the author in Ref.~\cite{dutra_parameterizations_2024}, reducing the number of decision variables and complexity of the estimator.
VI uses a different paradigm than the FEM for performing ML.
The FEM uses a filter to approximate the prior predictive distribution and, assuming that distribution is correct, computes the likelihood.
In VI, the estimation problem is augmented with the smoothing posterior distribution of the states, and a lower bound on the likelihood function is optimized instead.
The approximation error between the candidate posterior and the true one is encoded in the objective function.

In this paper, the application of variational system identification is demonstrated for estimating aircraft parameters from flight-test data, an important use case of practical interest \cite{grauer_introduction_2023}.
This is also the first application of the method in non-simulated data.
Our goal is to show that the estimator's usefulness and its results as a drop-in replacement for the FEM and OEM.
The results show that it is as good as, if not better than, those methods.
In particular, convergence is attained with null initial guesses for all system parameters, in all examples tested in this paper.

\section{Problem Formulation}
\label{sec:problem}
The goal of ML estimators for continuous-time dynamical systems is to obtain the vector $\theta\in\reals^{\npar}$ of unknown parameters of a system modeled by the stochastic differential equation (SDE)
\begin{equation}
  \label{eq:sde}
  \dd x(t) = f\!\big(x(t), u(t), \theta\big) \,\dd t + G(\theta)\,\dd W(t),
\end{equation}
where $x\colon\reals\to\reals^\nx$ are the states of the system, $u\colon \reals\to\reals^\ninp$ are the known external inputs, and the model functions $f$ and $G$ are given.
The SDE formulation is a more formal mathematical representation of the differential equation 
\begin{align}
  \label{eq:ode}
  \dot x(t) &= f\big(x(t), u(t), \theta\big) + G(\theta) w(t),
\end{align}
where $w(t)$ is the weak derivative of the Wiener process $W(t)$, i.e., white Gaussian noise with spectral density
\begin{equation}
  \label{eq:wiener_deriv}
    \E\left[w(t)w(\tau)\trans\right]=\ident \delta(t-\tau).
\end{equation}
We consider a continuous--discrete formulation where the states are indirectly measured at discrete time instants through the output vector
\begin{align}
  \label{eq:meas}
  y_k& = h\big(x(kT), u(kT), \theta\big) + v_k,
  \qquad
  k=0, 1, \dots, N;
\end{align}
where $k$ is the sample index, $v_k$ is the measurement noise, $T$ is the sampling period, and the output function $h$ is given.
The measurement noise is assumed to be a zero-mean, white, Gaussian noise process, independent from the process noise, with constant covariance $R(\theta)$:
\begin{equation}
  \label{eq:meas_cov}
  \E\left[v_iv_j\trans\right] = R(\theta)\delta_{ij}, \qquad i,j=0,1,\dotsc,N.
\end{equation}

To apply the estimator, the SDE is discretized using a method such as the Euler--Maruyama scheme, chosen here for its simplicity.
Given a prior distribution for the initial state $x_0$, the joint density of the discretized state path $x_0,x_1,\dotsc,x_N$, written in shorthand as $x_{0:N}$, is
\begin{equation}
  \label{eq:state_path_dens}
  p_\theta(x_{0:N}) = p_\theta(x_0) \prod_{k=1}^N p_\theta(x_{k}|x_{k-1}),
\end{equation}
where $p_\theta(x_0)$ is the initial-state prior density, and the state transition density under the Euler--Maruyama scheme is the multivariate normal with the following parameters:
\begin{align}
  p_\theta(x_{k}|x_{k-1}) &= 
  \normald\!\big(x_k; F(x_{k-1}, u_{k-1}, \theta), Q(\theta)\big), 
  \\
  F(x_{k-1}, u\kprev, \theta) &= x_k + Tf(x_{k-1}, u\kprev, \theta), 
  \\
  Q(\theta) &= T G(\theta)G(\theta)\trans.
\end{align}
Similarly, Eqs.~\eqref{eq:meas}--\eqref{eq:meas_cov} imply that
\begin{align}
  p_\theta(y_k|x_k) &= \normald\!\big(y_k; h(x_{k-1}, u_{k-1}, \theta), R(\theta)\big), \\
  \label{eq:meas_like_decomp}
  p_\theta(y_{0:N}|x_{0:N}) &= \prod_{k=0}^N p_\theta(y_k|x_k),
\end{align}
laying the groundwork for applying the estimator.

\section{Variational Inference}
\label{sec:vi}
In this section, the variational inference formulation of the maximum likelihood problem is derived, in a manner similar to \cite{courts_variational_2021, dutra_parameterizations_2024}.
Starting from Bayes' rule,
\begin{equation}
  \label{eq:bayes}
  p_\theta(x\kpath|y\kpath) = \frac{p_\theta(x\kpath,y\kpath)}{p_\theta(y\kpath)},
\end{equation}
we have that for every $x\kpath$ in the support of the posterior, the likelihood $p_\theta(y\kpath)$ of a parameter vector $\theta$ is the ratio between the density of the complete data and the posterior:
\begin{equation}
  \label{eq:likelihood_ratio}
  p_\theta(y\kpath) = \frac{p_\theta(x\kpath,y\kpath)}{p_\theta(x\kpath|y\kpath)}.
\end{equation}
This equation is not of practical use at any single state path $x\kpath$, but can be used to bound the likelihood when evaluated at a population of paths.
For any density $q(x\kpath)$ whose support lies in the support of the posterior%
\footnote{Or, equivalently, the measure associated with $q$ is absolutely continuous with respect to the posterior probability measure.},
we have that
\begin{align}
  \log p_\theta(y\kpath) = \log\left(\frac{p_\theta(x\kpath,y\kpath)}{q(x\kpath)}\times
    \frac{q(x\kpath)}{p_\theta(x\kpath|y\kpath)}\right) = 
  \log\left(\frac{p_\theta(x\kpath,y\kpath)}{q(x\kpath)}\right) + 
  \log\left(\frac{q(x\kpath)}{p_\theta(x\kpath|y\kpath)}\right).
\end{align}
Taking the expectation with respect to $q$, we have that
\begin{align}
  \label{eq:like_bound}
  \log p_\theta(y\kpath) &=
  \int\log\left(\frac{p_\theta(x\kpath,y\kpath)}{q(x\kpath)}\right)q(x\kpath)\,\dd x\kpath  + 
  \int\log\left(\frac{q(x\kpath)}{p_\theta(x\kpath|y\kpath)}\right)q(x\kpath)\,\dd x\kpath.
\end{align}

The last term on the right-hand side of Eq.~\eqref{eq:like_bound} is the Kullback--Leibler divergence (KLD) of $q$ from the posterior, which is always non-negative
\begin{align}
  \label{eq:kld}
  \kld[q(x\kpath) \Vert p_\theta(x\kpath | y\kpath)] = 
  \int\log\left(\frac{q(x\kpath)}{p_\theta(x\kpath|y\kpath)}\right)q(x\kpath)\,\dd x\kpath\geq 0.
\end{align}
The KLD, also known as relative entropy, quantifies the difference between two probability distributions, being zero if and only if they are equal almost everywhere.
This means that the fist term on the right-hand side of Eq.~\eqref{eq:like_bound} is a lower bound on the likelihood, or model evidence, which we will denote the evidence lower bound (ELBO),
\begin{align}
  \label{eq:elbo}
  \elbo(\theta, q) := \int\log\left(\frac{p_\theta(x\kpath,y\kpath)}{q(x\kpath)}\right)q(x\kpath)\,\dd x\kpath.
\end{align}
Substituting Eqs.~\eqref{eq:kld}--\eqref{eq:elbo} into Eq.~\eqref{eq:like_bound}, it is evident that the bound gets tighter as $q$ approaches the posterior, as quantified by the KLD:
\begin{align}
  \label{eq:like_bound_simple}
  \log p_\theta(y\kpath) &=
  \elbo(\theta, q) + \kld[q(x\kpath) \Vert p_\theta(x\kpath | y\kpath)] \geq \elbo(\theta, q).
\end{align}

This lays out the basis of variational inference for maximum likelihood estimation: by searching for both the parameter vector $\theta$ and the density $q$ which maximize the ELBO $\elbo(\theta, q)$, we are simultaneously maximizing the likelihood function and minimizing the divergence from $q$ to the true posterior.
The density $q$, called the assumed density, should be chosen from a family of distributions $\qspace$ that is general enough to yield tight bounds, but still be tractable enough to compute and optimize the ELBO.
The ML estimation problem using VI is then
\begin{equation}
  \label{eq:vi_ml}
  \operatorname*{maximize}_{q\in\qspace,\theta\in\reals^\npar} \elbo(\theta, q).
\end{equation}
In situations where the true posterior is contained in $\qspace$, this formulation is equivalent to ML estimation.
When the likelihood function is intractable to compute, as it is in parameter estimation of nonlinear systems, VI with its assured lower bound offers stronger theoretical guarantees than when approximations like the extended Kalman filter or particle smoothers are used to compute and maximize the likelihood \cite{kristensen_parameter_2004, schon_system_2011}.

VI can also be understood as a regularized generalization of maximum \emph{a posteriori} (MAP) estimation over a population of points \cite[see][Sec.~3.2]{dutra_parameterizations_2024}.
Eq.~\eqref{eq:elbo} can be split into the expected complete-data log-density and the differential entropy of the assumed density $q$:
\begin{align}
  \elbo(\theta, q) &= \int\log p_\theta(x\kpath,y\kpath) q(x\kpath)\,\dd x\kpath 
  -\int\log q(x\kpath)q(x\kpath)\,\dd x\kpath. \\
  \label{eq:elbo_decomp}
  &= \int\log p_\theta(x\kpath,y\kpath) q(x\kpath)\,\dd x\kpath +
  \entro_q[X\kpath].
\end{align}
In MAP estimation, a single state path is chosen as the most probable, given the observations.
In VI, a population of paths, defined by $q$, is chosen to take into account the uncertainty in the states.
The differential entropy of the population of paths $\entro_q[X\kpath]$, which increases as the population is spread out over a larger region, is included to prevent the population from collapsing at a single point, the posterior mode.

\section{Variational System Identification}
As discussed in the previous section, to apply VI for solving a ML estimation problem, we need to define both the underlying dynamic system and noise model, as in Sec.~\ref{sec:problem}, as well as the search space $\qspace$ of the assumed density $q$.
For the problem defined in Sec.~\ref{sec:problem}, a key property for making the underlying optimization problem sparse is to choose $q$ from a space of densities that can be represented by a factor graph with the same topology as that of the true posterior.
This amounts to choosing $q$ under which the process $x_k$ has the Markov property \cite[see][Sec.~5]{dutra_parameterizations_2024}.

For tractability in computing the ELBO in Eq.~\eqref{eq:elbo_decomp}, we also choose the assumed density $q$ to be multivariate normal, as it has many tractable methods for computing expectations with respect to it.
The differential entropy of the multivariate normal has a simple closed-form expression.
As shown in Ref.~\cite[Sec.~5]{dutra_parameterizations_2024}, a multivariate normal distribution $q(x_{0:N})$ with the Markov property can be represented as
\begin{align}
  q(x_{0:N}) &= q(x_0)\prod_{k=1}^N q(x_k|x_{k-1}),
\end{align}
where each term is given by
\begin{align}
  \label{eq:gvi_x0_q_gen}
  q(x_0) &= \normald(x_0; \mu_0, \Sigma_0),
  &
  q(x_k|x_{k-1}) &= \normald(x_0; \mu_k+\Sigma_{k,k-1}\Sigma_{k-1}\inv(x_{k-1}-\mu_{k-1}), \Sigma_{k|k-1}).
\end{align}

In linear systems and nonlinear systems operating at small deviations form an equilibrium point, as is often the case in aircraft system identification, the posterior convariances converge to steady-state values when the data record is long enough.
Under this simplification, the covariances for all $k=0,\dotsc,N$ are given by
\begin{align}
  \Sigma_{k|k-1}&=\Sigma\cond,
  &
  \Sigma_{k,k-1}&=\Sigma\cross,
  &
  \Sigma_k& = \Sigma\cond + \Sigma\cross\Sigma\marg\inv\Sigma\cross\trans,
\end{align}
and $\mu_{0:N}$, $\Sigma\cond\succ 0$, and $\Sigma\cross$ are the parameters of $q$.
This is the steady-state parameterization of Ref.~\cite{dutra_parameterizations_2024}, which is an analog of the FEM using a steady-state Kalman filter.
We also note that this representation has a very simple differential entropy, as the chain rule for differential entropy gives
\begin{equation}
  \label{eq:entro}
  \entro_q[X\kpath] = \entro_q[X_0] + \sum_{k=1}^N \entro_q[X_k|X\kprev] = \tfrac12\log\det\Sigma\marg + \tfrac N2\log\det\Sigma\cond + \operatorname{const},
\end{equation}
where const denotes constant terms which do not change the location of maxima.

With the parameterization of the assumed density defined, we can return to the objective function defined in Eq.~\eqref{eq:elbo_decomp} and substitute Eqs.~\eqref{eq:state_path_dens} and~\eqref{eq:meas_like_decomp} to obtain
\begin{multline}
  \label{eq:elbo_reduced}
  \elbo(\theta, q) = 
  \int\log p_\theta(x_0) q(x_0)\,\dd x_0
  + \sum_{k=1}^N \int\log p_\theta(x_{k}|x_{k-1}) q(x_{k-1:k})\,\dd x_{k-1:k} \\
  + \sum_{k=0}^N \int\log p_\theta(y_{k}|x_{k}) q(x_{k})\,\dd x_{k}
  + \entro_q[X\kpath].
\end{multline}
Instead of having an expectation over a high-dimensional space, this form of the ELBO only requires expectation over the states at each time sample or at consecutive time-sample pairs.
This allows integration using sigma points, which is exact for linear systems and a reasonable approximation for nonlinear systems under small deviations from trim.

For simplicity, in this paper we use sigma points without the center one, which gives all remaining points equal weight.
If we let $x\sigp_k$ denote the sigma points associated with $q(x_k)$ or $q(x_{k-1:k})$, the objective function is then
\begin{multline}
  \elbo(\theta, q) = 
  \frac{1}{2\nx}\sum_{i=1}^{2\nx} p_\theta(x\sigp_0) 
  + \frac{1}{4\nx}\sum_{k=1}^N \sum_{i=1}^{4\nx} \log p_\theta(x\sigp_{k}|x\sigp_{k-1})
  + \frac{1}{2\nx}\sum_{k=0}^N \sum_{i=1}^{2\nx}\log p_\theta(y_{k}|x\sigp_{k}) \\
  + \tfrac12\log\det\Sigma\marg
  + \tfrac N2\log\det\Sigma\cond + \operatorname{const}.
\end{multline}
This function can then be maximized using standard nonlinear programming packages.
Note that $x\sigp_k$ depend on $q$ and its underlying parameters.
Changes in $\mu_{0:N}$ translate all sigma points by some amount, while increasing $\Sigma\marg$ increases their spread, decreasing the average log-density but increasing the entropy.
The optimum balances these competing criteria.

\section{Results}
This section describes the application of the variational system identification to real flight-test data.
Freely accessible data from Refs.~\cite{jategaonkar_flight_2015, morelli_aircraft_2016} was used to demonstrate the method.
The data are of a HFB-320 Hansa Jet, a VFW-Fokker 614, and a DHC-6 Twin Otter.
These represent a wide range of aircraft: a business jet, a jet airliner, and a STOL utility aircraft.
The models used in the textbook examples of Refs.~\cite{jategaonkar_flight_2015, morelli_aircraft_2016} were used in this paper.

The open source Python package \texttt{visid}, developed by the author and released\footnote{\url{https://github.com/dimasad/visid}} on Github, was used to implement the variational estimator.
The code for the experiments in this paper is also released\footnote{\url{https://github.com/dimasad/scitech-2025-code}} on Github as open source software.
The underlying optimization problem was solved with the open-source nonlinear program solvers in \texttt{scipy.optimize.minimize}, part of the SciPy package \cite{virtanen_scipy_2020}.
Automatic differentiation with JAX \cite{jax2018github} was used to compute the objective function's gradient and Hessian.
The initial guess of all optimization decision variables was zero, except for the HFB-320, which had the smoother mean $\mu_{0:N}$ initially set to the measurements of the states in $y_{0:N}$.

As argued in Ref.~\cite[Sec.~IV.B]{grauer_new_2015}, care must be taken when evaluating the results of the FEM, as the filter corrections can yield a very good match to the data.
The same applies to the variational method with respect to the smoother output.
Because of this, the optimal parameters $\theta\opt$ estimated by variational system identification were evaluated using the following criteria:
\begin{enumerate}
\item The output associated to the smoother mean, that is, $y_k\opt = h(\mu_k, u_k, \theta\opt)$.
\item The output of the free simulation using $\theta\opt$.
\item The one-step-ahead predictions of the steady-state Kalman filter associated to $\theta\opt$.
\item The difference between the smoother-mean finite differences, $\frac{\mu_{k+1}-\mu_k}{T}$ and the model drift function $f(\mu_k, u_k, \theta\opt)$.
\end{enumerate}
These quantities show the two things the estimator seeks to minimize: the output error and the equation error.
When two repetitions of the same maneuver are available, one dataset is used for estimation and another for validation.
If only one repetition is available, no validation is performed.
For validation, the VI method is run with the estimated model parameters $\theta\opt$ fixed, for estimation of the assumed density $q$ only (smoothing).

\begin{figure}
  \tikzsetnextfilename{hfb.tikz}%
  \tikzpicturedependsonfile{fig/hfb.tikz.tex}%
  \input{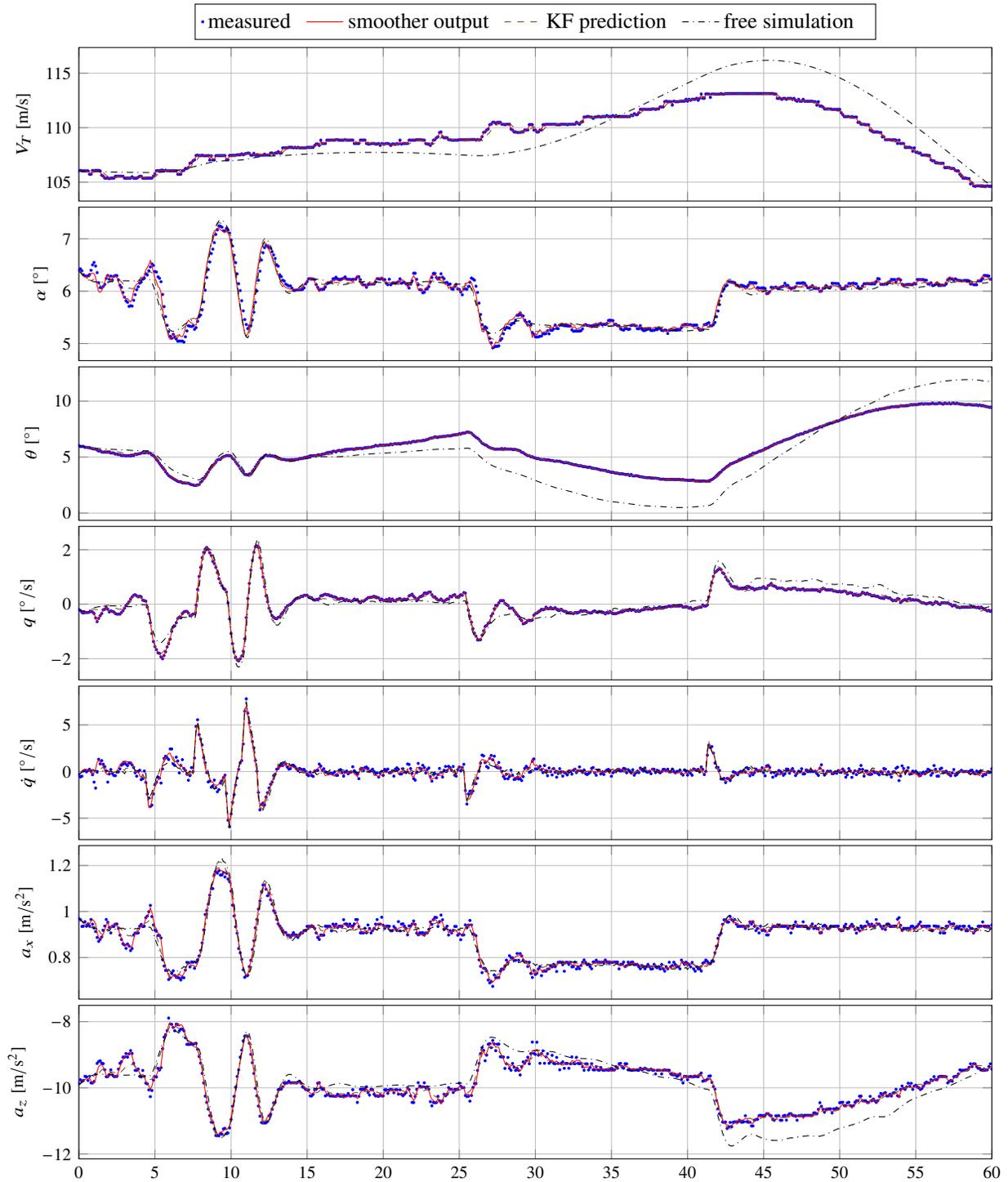}%

  \caption{
    Evaluation of model outputs for the optimal parameters estimated by variational system identification on the HFB-320 flight-test data.
    This is the same dataset that was used to estimate the model.
  }
  \label{fig:hfb_y}
\end{figure}

\begin{figure}[p]
  \tikzsetnextfilename{hfb-xdot.tikz}%
  \tikzpicturedependsonfile{fig/hfb-xdot.tikz.tex}%
  \input{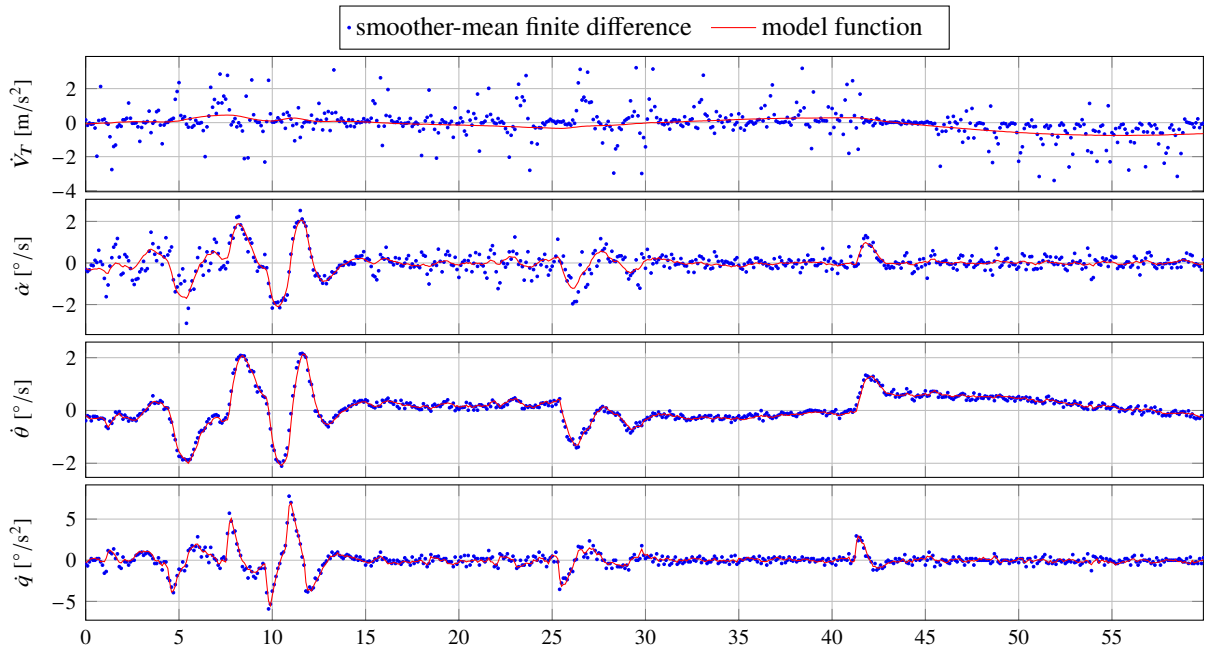}%

  \caption{
    Evaluation of state derivatives for the optimal parameters estimated by variational system identification on the HFB-320 flight-test data.
    This is the same dataset that was used to estimate the model.
  }
  \label{fig:hfb-xdot}
\end{figure}

\begin{figure}[p]
  \tikzsetnextfilename{totter_lat-xdot.tikz}%
  \tikzpicturedependsonfile{fig/totter_lat-xdot.tikz.tex}%
  \input{fig/totter_lat-xdot.tikz.tex}%

  \caption{
    Evaluation of state derivatives for the optimal parameters estimated by variational system identification on the DHC-6 lateral-directional motion flight-test data.
    This is the same record that was used to estimate the model.
  }
  \label{fig:totter_lat-xdot}
\end{figure}

\begin{figure}[p]
  \tikzsetnextfilename{totter_lat.tikz}%
  \tikzpicturedependsonfile{fig/totter_lat.tikz.tex}%
  \input{fig/totter_lat.tikz.tex}%

  \caption{
    Evaluation of model outputs for the optimal parameters estimated by variational system identification on the DHC-6 lateral-directional motion flight-test data.
    This is the same record that was used to estimate the model.
  }
  \label{fig:totter_lat_y}
\end{figure}

\begin{figure}[p]
  \tikzsetnextfilename{totter_lat_val-xdot.tikz}%
  \tikzpicturedependsonfile{fig/totter_lat_val-xdot.tikz.tex}%
  \input{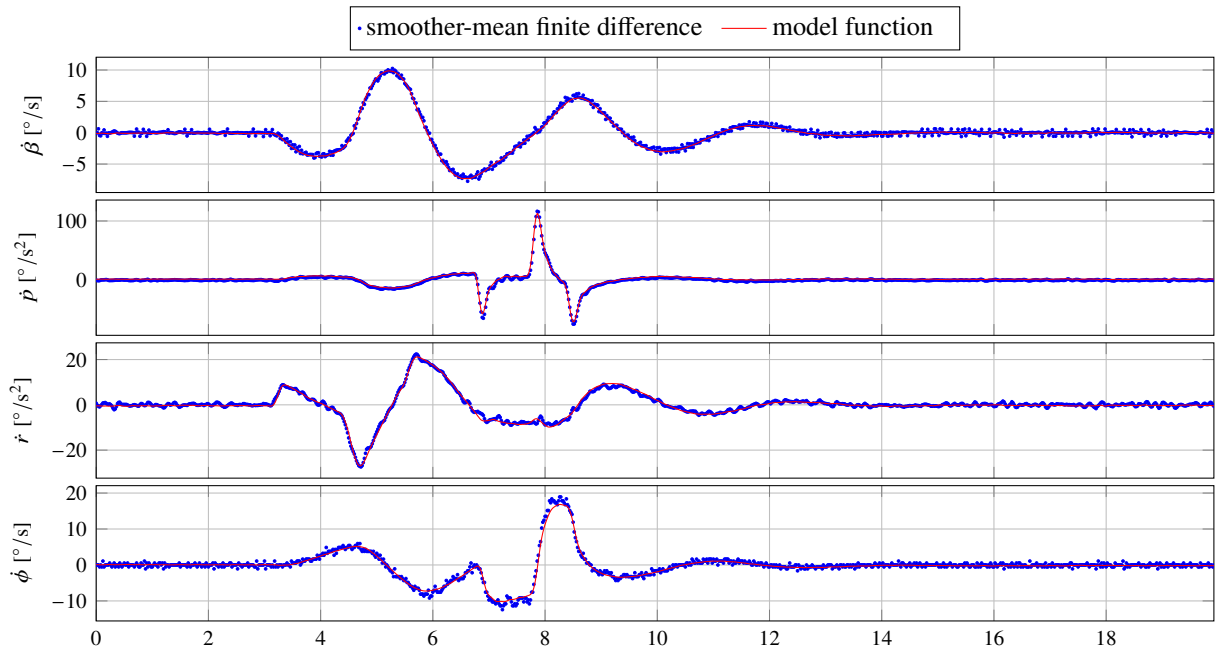}%

  \caption{
    Evaluation of state derivatives for the optimal parameters estimated by variational system identification on the DHC-6 lateral-directional motion validation flight-test data.
    This is a different record than the one used to estimate the model.
  }
  \label{fig:totter_lat_val-xdot}
\end{figure}

\begin{figure}[p]
  \tikzsetnextfilename{totter_lat_val.tikz}%
  \tikzpicturedependsonfile{fig/totter_lat_val.tikz.tex}%
  \input{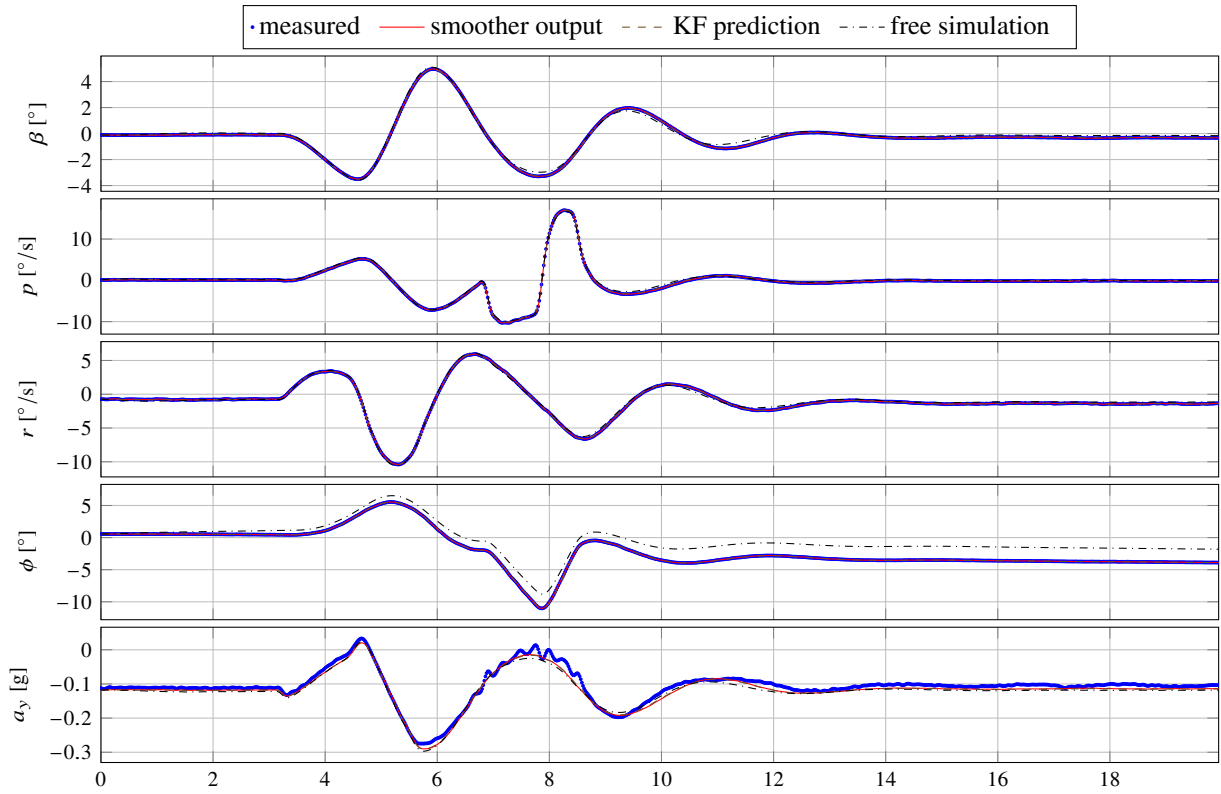}%

  \caption{
    Evaluation of model outputs for the optimal parameters estimated by variational system identification on the DHC-6 lateral-directional motion valitation flight-test data.
    This is a different record than the one used to estimate the model.
  }
  \label{fig:totter_lat_val_y}
\end{figure}

\begin{figure}[p]
  \tikzsetnextfilename{totter_sp.tikz}%
  \tikzpicturedependsonfile{fig/totter_sp.tikz.tex}%
  \input{fig/totter_sp.tikz.tex}%

  \caption{
    Evaluation of model outputs for the optimal parameters estimated by variational system identification on the DHC-6 short period motion flight-test data.
    This is the same record that was used to estimate the model.
  }
  \label{fig:totter_sp_y}
\end{figure}

\subsection{Results for the HFB-320}
\label{sec:hfb}

This example corresponds to an elevator 3-2-1-1 input followed by a pulse on a HFB-320 of the \emph{Deutsches Zentrum für Luft- und Raumfahrt} (DLR).
The application of both the OEM and FEM to this dataset is reported in Ref.~\cite[Sec.~5.11.2]{jategaonkar_flight_2015}.
The collocation-based OEM was also applied to this dataset in Ref.~\cite[Sec.~III.A]{dutra_collocation-based_2019}.
A nonlinear longitudinal model with 4~states, 7~outputs and 2~inputs was fit to the data.
Please refer to Refs.~\cite{jategaonkar_flight_2015, dutra_collocation-based_2019} or the accompanying software of this paper and its references for the model equations.


The model fit to the estimation data is shown in Figs.~\ref{fig:hfb_y}--\ref{fig:hfb-xdot}.
A single dataset was provided, there is no independent validation dataset.
As noted in Ref.~\cite{jategaonkar_flight_2015}, there is considerable atmospheric turbulence in this record, as can be seen in the measurements of $\alpha$ and $a_z$ in Fig.~\ref{fig:hfb_y}.
The free simulation shows a similar fit than the OEM in Refs.~\cite{jategaonkar_flight_2015, dutra_collocation-based_2019}, with most error in $V_T$ and $\theta$ where the low damping and frequency of the phugoid accumulates perturbations due to turbulence.

The smoother output and KF predictor are very close to the FEM results in Ref.~\cite{jategaonkar_flight_2015}.
A closer inspection of Fig.~\ref{fig:hfb-xdot} shows that the derivative of the smoothed state mean is very noisy for $V_T$ and $\alpha$.
This is likely a result of excessive smoother corrections due to overconfidence in the measured values.
The measurements of $V_T$, $a_x$, and $a_z$ have visible quantization levels and the estimated measurement noise standard deviations are well below this interval.
No bounds were placed on $R$ or $Q$ in this example, however, for comparison with the other references.

\subsection{Results for the DHC-6 Lateral-Directional Motion}
This example corresponds to aileron and rudder multi-step inputs on a DHC-6 Twin Otter of the NASA Glenn Research Center.
The application of both the OEM and EEM to this dataset is reported in Ref.~\cite[Examples~5.1, 6.1, and~7.1]{morelli_aircraft_2016}.
A linear lateral directional model with 4~states, 5~outputs and 2~inputs was fit to the data, parameterized by the nondimentional stability and control derivatives.
Please refer to Ref.~\cite{morelli_aircraft_2016} or the accompanying software of this paper and its references for the model equations.

The model fit to the estimation data is shown in  Figs.~\ref{fig:totter_lat-xdot}--\ref{fig:totter_lat_val_y}.
There is little turbulence in the records and an excelent match is obtained for the estimation (Figs.~\ref{fig:totter_lat-xdot}--\ref{fig:totter_lat_y}) and validation (Figs.~\ref{fig:totter_lat_val-xdot}--\ref{fig:totter_lat_val_y}) datasets.
The divergence of the roll angle $\phi$ in the free simulation of the validation dataset is attributed to different values of the bias parameter in that record and the accumulation of error in the slow spiral mode.

Some quantization effects can be noticed in the first and last plots of Figs.~\ref{fig:totter_lat-xdot} and~\ref{fig:totter_lat_val-xdot}, but the signal to noise ratio is much larger than in the example of Sec.~\ref{sec:hfb} due to higher resolution.
The smoother is overcorrecting to $\beta$, $p$, and $r$, but by a small amount.
No bounds were placed on $R$ or $Q$ in this example as well, for comparison with the other references.

\begin{figure}[t]
  \tikzsetnextfilename{totter_sp-xdot.tikz}%
  \tikzpicturedependsonfile{fig/totter_sp-xdot.tikz.tex}%
  \input{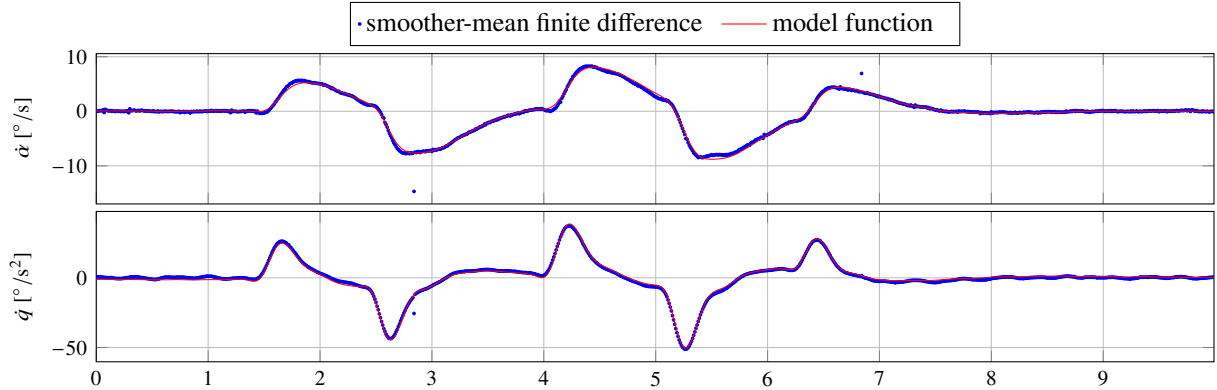}%

  \caption{
    Evaluation of state derivatives for the optimal parameters estimated by variational system identification on the DHC-6 short period motion flight-test data.
    This is the same record that was used to estimate the model.
  }
  \label{fig:totter_sp-xdot}
\end{figure}

\subsection{Results for the DHC-6 short period motion}
This example corresponds to an elevator multi-step input on a DHC-6 Twin Otter of the NASA Glenn Research Center.
The application of the OEM to this dataset is reported in Ref.~\cite[Example~6.3]{morelli_aircraft_2016}.
A linear short period motion model with 2~states, 3~outputs and 1~input was fit to the data, parameterized by the dimentional stability and control derivatives.
Please refer to Ref.~\cite{morelli_aircraft_2016} or the accompanying software of this paper and its references for the model equations.

The model fit to the estimation data is shown in  Figs.~\ref{fig:totter_sp_y}--\ref{fig:totter_sp-xdot}.
There is little turbulence in the records and an excelent match is obtained for the vertical acceleration $a_z$ and the pitch rate $q$.
A low-frequency error is present on the angle of attack $\alpha$ due to unmodelled phugoid dynamics.
A single dataset was provided, so there is no independent validation dataset.

\begin{figure}[t]
  \tikzsetnextfilename{attas_sp-xdot.tikz}%
  \tikzpicturedependsonfile{fig/attas_sp-xdot.tikz.tex}%
  \input{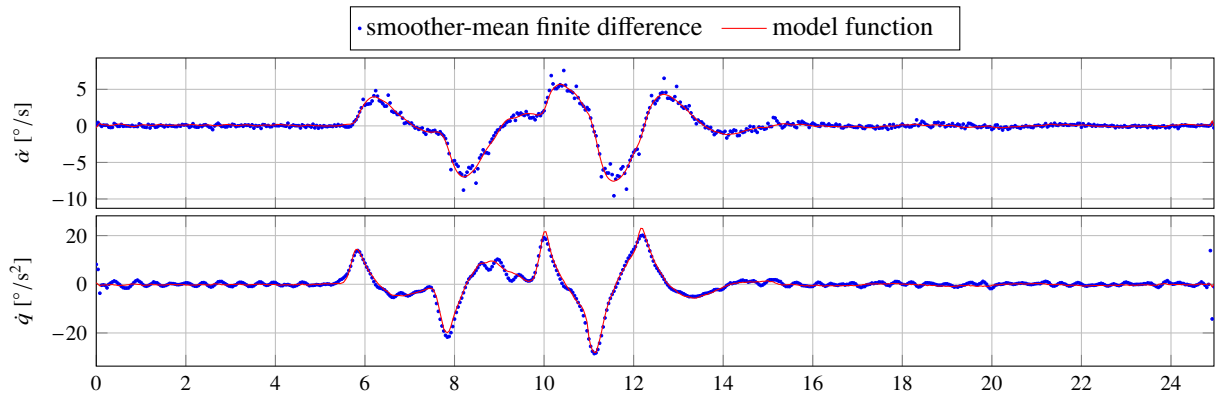}%

  \caption{
    Evaluation of state derivatives for the optimal parameters estimated by variational system identification on the VFW-Fokker 614 short period motion flight-test data.
    This is the same record that was used to estimate the model.
  }
  \label{fig:attas_sp-xdot}
\end{figure}

\begin{figure}
  \tikzsetnextfilename{attas_sp.tikz}%
  \tikzpicturedependsonfile{fig/attas_sp.tikz.tex}%
  \input{fig/attas_sp.tikz.tex}%

  \caption{
    Evaluation of model outputs for the optimal parameters estimated by variational system identification on the VFW-Fokker 614 short period motion flight-test data.
    This is the same record that was used to estimate the model.
  }
  \label{fig:attas_sp_y}
\end{figure}

\begin{figure}
  \tikzsetnextfilename{attas_sp_val-xdot.tikz}%
  \tikzpicturedependsonfile{fig/attas_sp_val-xdot.tikz.tex}%
  \input{fig/attas_sp_val-xdot.tikz.tex}%

  \caption{
    Evaluation of state derivatives for the optimal parameters estimated by variational system identification on the VFW-Fokker 614 short period motion validation flight-test data.
    This is a different record than the one used to estimate the model.
  }
  \label{fig:attas_sp_val-xdot}
\end{figure}

\begin{figure}
  \tikzsetnextfilename{attas_sp_val.tikz}%
  \tikzpicturedependsonfile{fig/attas_sp_val.tikz.tex}%
  \input{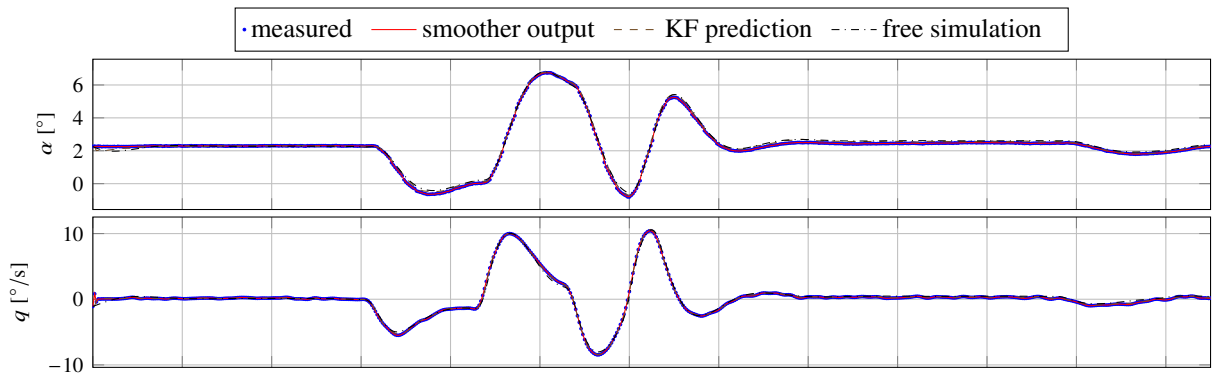}%

  \caption{
    Evaluation of model outputs for the optimal parameters estimated by variational system identification on the VFW-Fokker 614 short period motion validation flight-test data.
    This is a different record than the one used to estimate the model.
  }
  \label{fig:attas_sp_val_y}
\end{figure}

\subsection{Results for the VFW-Fokker 614 Short Period Motion}
This example corresponds to an elevator 3-2-1-1 input on a VFW-Fokker 614 of the DLR.
The application of the OEM and nonlinear filters to this dataset is reported in Ref.~\cite[Sec.~7.5.2]{jategaonkar_flight_2015}.
A linear longitudinal model with 2~states, 2~outputs and 1~input was fit to the data, parameterized by the dimensional stability and control derivatives.
Please refer to Ref.~\cite{jategaonkar_flight_2015} or the accompanying software of this paper and its references for the model equations.

The model fit to the estimation data is shown in  Figs.~\ref{fig:attas_sp-xdot}--\ref{fig:attas_sp_val_y}.
There is little turbulence in the records and an excelent match is obtained for the estimation (Figs.~\ref{fig:attas_sp-xdot}--\ref{fig:attas_sp_y}) and validation (Figs.~\ref{fig:attas_sp_val-xdot}--\ref{fig:attas_sp_val_y}) datasets.
The free simulation of the estimated model is very close to the results of the OEM in Ref.~\cite{jategaonkar_flight_2015}.

\section{Conclusion and Future Work}
This paper showed the application of variational system identification for a wide range of models and aircraft.
For flights in turbulence, like shown in Sec.~\ref{sec:hfb}, the results of the method are very similar to those of the FEM.
A major practical advantage of variational system identification over the OEM and FEM, however, is that the estimator converges to the optimum, for all examples shown, with the initial guess of all system parameters set to zero.

Beyond the applications shown here, variational system identification has the potential to enable treatment of both process and measurement noise in problems where the FEM or OEM are not currently applicable.
As shown in \cite{dutra_parameterizations_2024}, variational system identification can be used for processing long records of data, to estimate parameters from models with a large number of states, and to estimate models for which good initial parameter estimates are not available.
The computation of the objective function and its derivatives is embarrassingly parallel and can benefit from great speedups when computed in modern graphical processing units (GPUs).
The strong theoretical support of VI also makes it a great candidate for global nonlinear modeling and treatment of non-Gaussian noise.

\section*{Acknowledgments}
This research was made possible by the NASA Established Program to Stimulate Competitive Research, Grant \#~80NSSC20M0055.
Computational resources were provided by the WVU Research Computing Thorny Flat and Dolly Sods HPC clusters, which are funded in part by NSF OAC-1726534 and NSF OAC-2117575.
\bibliography{bib}

\end{document}